\documentclass{article}

\usepackage{mathrsfs}
\usepackage{amsfonts}
\usepackage{amsmath}
\usepackage{amssymb}
\usepackage{graphicx}

\title{The Topology of Branching Universes}
\author{Gordon McCabe}

\begin{document}

\maketitle

\begin{abstract}

The purpose of this paper is to survey the possible topologies of
branching space-times, and, in particular, to refute the popular
notion in the literature that a branching space-time requires a
non-Hausdorff topology.

\end{abstract}

\section{Branching MWI space-times}

Two basic types of branching universe have been suggested in
modern mathematical physics: universes which branch in a style
befitting the `many-worlds' interpretation (MWI) of quantum
theory; and universes which bifurcate to implement a
topology-change cobordism.

This section, concerned with the former type of branching
universe, is not intended as a defence or analysis of the
many-worlds interpretation. It is not even the intention to argue
that the many-worlds interpretation genuinely entails such a
branching model of space-time. Rather, the intention is to analyse
the topological implications of such a branching model of
space-time, and to question some of the assumptions normally found
in the literature.

The many-worlds interpretation holds that if a system is prepared
into a state which is a `superposition' $\Psi = \sum_i a_i v_i$
with respect to some quantity $A$ whose eigenvectors provide a
basis $\{ v_i \}$ of the quantum state space, then when that
system undergoes a measurement-like interaction, the universe
splits into multiple branches, one for each basis vector $v_i$.
Each different branch records a definite value for the quantity
$A$, corresponding to the eigenvector $v_i$.

This branching can be thought of as either a local or global
process. If the branching is considered to be a global process,
then the branches correspond to multiple regions of
four-dimensional space-time which emanate from a common
three-dimensional hypersurface, (see Figure 1). Each such
branching hypersurface has a non-unique future.

\begin{figure}
\centering
\includegraphics[scale = 0.5]{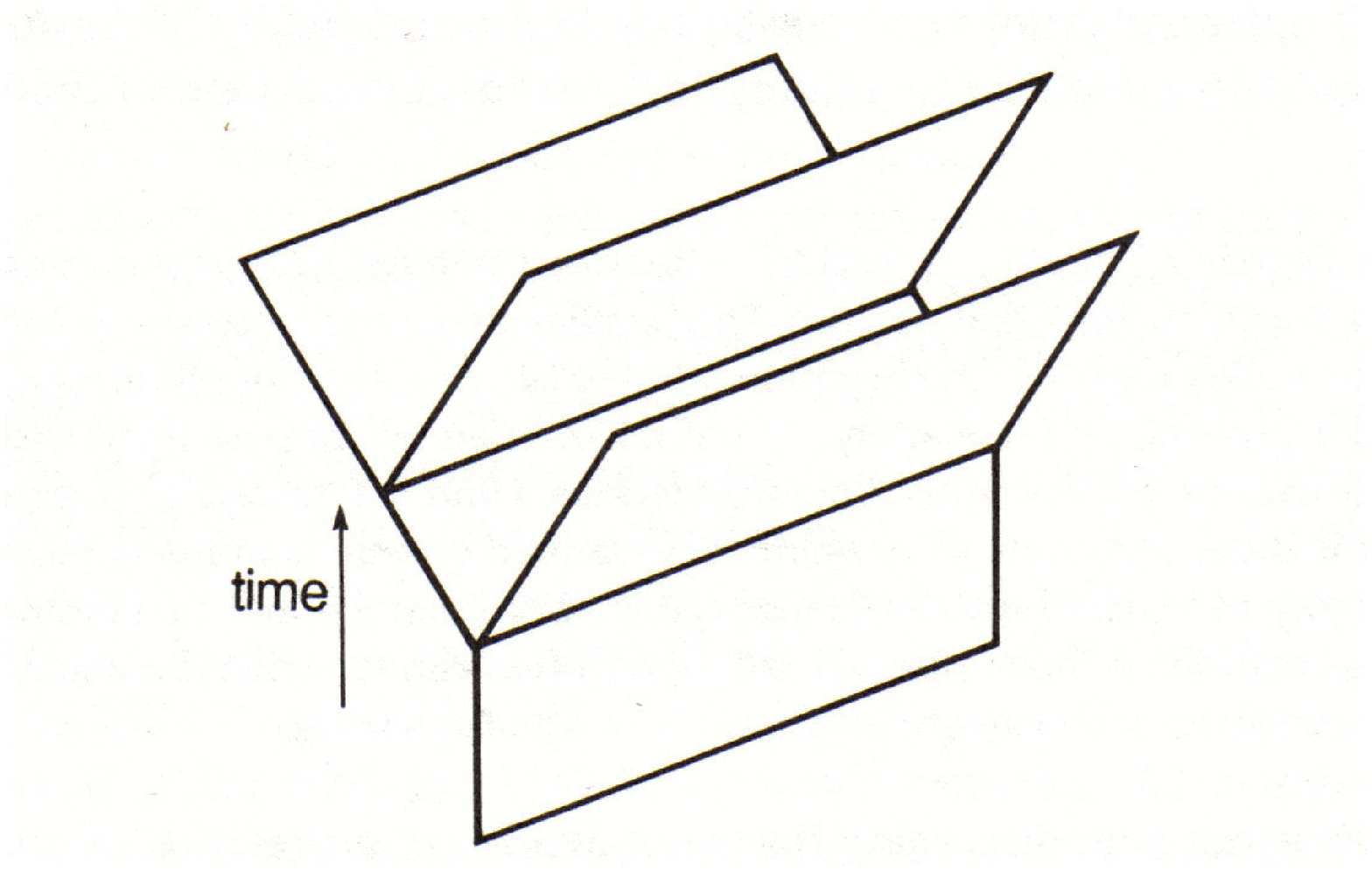}
\caption{Globally bifurcating space-time, from Earman 1986, p225}
\end{figure}

The notion that the entire universe branches in this style every
time there is a measurement-like interaction, renders such
branching a highly non-local process, and tacitly supposes that
there is a unique global time coordinate for the universe.
Treating a measurement-like interaction as a point event in
space-time, there will be many spacelike hypersurfaces which pass
through that point; the selection of only one of these as the
branching hypersurface requires one to accept that there is a
preferential time coordinate for the universe. To avoid these
difficulties, one can suggest that the universe only branches
locally as the result of a measurement-like interaction. To be
specific, one can suggest that the future light-cone of the
interaction event has multiple branches, one for each possible
outcome of the interaction. If one imagines such a universe as a
two-dimensional sheet, then the image is one in which there are
numerous pockets in the sheet, formed by the multiple branches of
future light cones. Penrose has drawn just such an image of a
branching MWI universe, (see Figure 2).

\begin{figure}
\centering
\includegraphics[scale = 0.4]{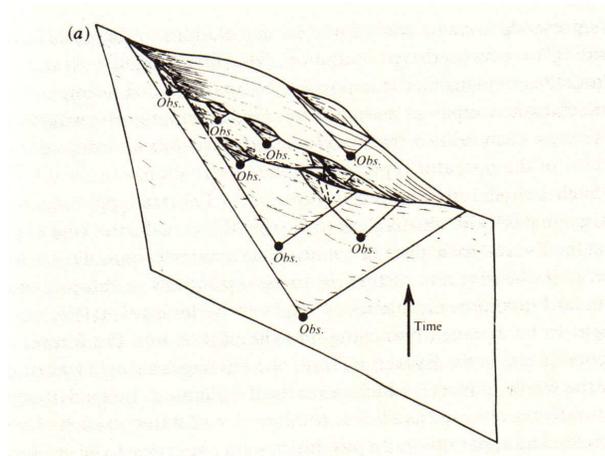}
\caption{Bifurcating light cones, from Penrose 1979, p593}
\end{figure}

It has been repeatedly claimed, by different authors, for over two
decades, that such a model of space-time entails a non-Hausdorff
topology, and therefore cannot be a topological manifold. In 1979
Penrose suggests that such branching space-times ``could be
legitimate mathematical objects, e.g. four-dimensional Lorentzian
manifolds subject to Einstein's field equations (say), but where
the Hausdorff condition is dropped," (1979, p594). In 2004,
Penrose repeats this assertion, suggesting that ``non-Hausdorff
manifolds can `branch'," (2004, p222). In the interim, Matt Visser
has echoed Penrose's assertion, stating that one ``uses
non-Hausdorff manifolds to describe `train track' geometries where
the same present has two or more futures," (2003, p164), and
claiming that ``a non-Hausdorff manifold has the bizarre property
that the dimensionality of the manifold is not necessarily equal
to the dimensionality of the coordinate patches. From a
physicist's perspective, this idea has been explored somewhat by
Penrose," (1993, p12). Lee Smolin, dealing with a `minimalist
wormhole' obtained by identifying a pair of points in a manifold,
states that ``of course, we loose the Hausdorff property," (1994,
p2).

Let us review the pertinent definitions to gain some clarity. A
topological manifold is a topological space which is:

\begin{enumerate}

\item Locally Euclidean. i.e each point $p$ possesses an open
neighbourhood $U$ which is homeomorphic with an open subset of
some $n$-dimensional Euclidean space $\mathbb{R}^n$. The
homeomorphism $\phi:U \rightarrow \mathbb{R}^n$ is called a
coordinate chart, and $n$ is the local dimension.

\item A Hausdorff topological space. i.e each pair of points $p,q$
possess neighbourhoods, $p \in U$ and $q \in V$, which are
disjoint from each other, $U \cap V = \emptyset$.

\end{enumerate}

A connected locally Euclidean topological space must be of
constant local dimension.

Manifolds of various type, such as smooth manifolds or
piecewise-linear manifolds, are topological manifolds which
satisfy additional conditions concerning the transformations from
one coordinate chart to another, wherever those charts overlap.

Penrose and Visser are suggesting that a branching space-time
corresponds to a non-Hausdorff topological space which is still
locally Euclidean, of dimension 4, and equipped with a smooth
Lorentzian metric. As Visser states, ``normally manifolds are
assumed Hausdorff by definition. Relaxing this condition permits
the existence of `branched' manifolds," (1996, p250). Whilst this
is true, one can equally represent branching space-times with
topological spaces which are Hausdorff, but not locally Euclidean
about every point. Moreover, the diagrams used to represent
branching space-times correspond to the latter case, and this fact
has been neglected in the literature.

Let us consider a simple example of the distinction between these
two types of branching topology. Consider first the example given
by Visser of a one-dimensional manifold that branches in two:

Begin with the real line $\mathbb{R}$, remove the half-closed
interval $I = [0,\infty)$, and replace it with two copies $I_1 =
[0_1,\infty)$ and $I_2 = [0_2,\infty)$ by taking the union $M =
(-\infty, 0) \cup [0_1,\infty) \cup [0_2,\infty)$. Define a basis
for the topology on $M$ by requiring that any open subset of
$(-\infty, 0) \cup [0_1,\infty)$ and any open subset of $(-\infty,
0) \cup [0_2,\infty)$ is an open subset of $M$. This means that
the neighbourhood base of $0_1$ is provided by the open subsets
$(-\epsilon, 0) \cup [0_1,\epsilon)$, for $\epsilon > 0$, and the
neighbourhood base of $0_2$ is provided by the open subsets
$(-\epsilon, 0) \cup [0_2,\epsilon)$. As a consequence $0_1$ and
$0_2$ do not possess disjoint neighbourhoods, and the topological
space is non-Hausdorff, but still locally Euclidean of dimension 1
about every point. This example of a 1-dimensional non-Hausdorff
space is also that given by Hawking and Ellis (1973, p13-14), and
Geroch's example of a non-Hausdorff manifold is essentially just a
2-dimensional analogue (1971, p100).

\begin{figure}
\centering
\includegraphics[scale = 0.5]{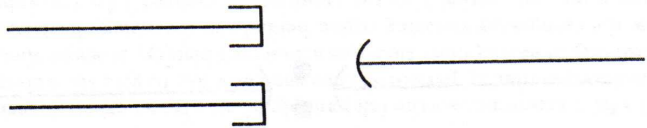}
\caption{Non-Hausdorff but locally Euclidean, from Visser 1996,
p252}
\end{figure}

Visser provides two diagrams to represent this non-Hausdorff
space. The first diagram, (see Figure 3), clearly represents the
fact that $0_1$ and $0_2$ are distinct points, but the second
(1996, p253) treats both points as the vertex of a $Y$-shape, and
is potentially misleading. Whilst the first diagram is correct,
the second, (see Figure 4), is conventionally taken to represent
the following \emph{Hausdorff} space, in which $0_1$ and $0_2$ are
identified:

Begin with the real line $\mathbb{R}$, remove the half-closed
interval $I = [0,\infty)$, and replace it with two copies $I_1 =
[0_1,\infty)$ and $I_2 = [0_2,\infty)$ by (i) taking the union $M
= (-\infty, 0) \cup [0_1,\infty) \cup [0_2,\infty)$; (ii) defining
an equivalence relationship $R$ which is such that $0_1 \sim 0_2$;
(iii) taking the quotient topological space $Q = M \backslash R$.
The quotient space $Q$ consists of the set of equivalence classes
in $M$, where the only equivalence class containing more than one
point is $[0] = \{0_1,0_2\}$. The projection mapping $\pi:M
\rightarrow Q$ is such that $\pi(0_1) = \pi(0_2) = [0]$. The
quotient topology is defined to be the largest topology one can
bestow on $Q$ which still permits the projection mapping to be
continuous. Whilst there are potentially many topologies on $Q$
such that the inverse image $\pi^{-1}(W)$ of every open subset $W
\subset Q$ is an open subset of $M$, the quotient topology is the
strongest of these. A subset $W \subset Q$ is defined to be open
if and only if the inverse image $\pi^{-1}(W)$ is an open subset
of $M$.\footnote{Note, this does not mean that the image of every
open subset of $M$ projects onto an open subset of $Q$; there may
be open subsets $U \subset M$ which are such that
$\pi^{-1}(\pi(U))$ is not an open subset of $M$. The projection
mapping is said to be open if every open subset of $M$ does
project onto an open subset of $Q$.} This quotient construction
gives a branching topological space which genuinely corresponds
diagrammatically to a $Y$-shape. The quotient topological space is
Hausdorff, but not locally Euclidean about every point. There is
only one point associated with the branch, namely $[0]$, and this
point can be separated from any other point by disjoint
neighbourhoods. However, there is no neighbourhood of this point
which can be homeomorphically mapped to an open subset of
$\mathbb{R}$. Every neighbourhood of $[0]$ is $Y$-shaped, and such
a subset cannot be mapped to an open subset of $\mathbb{R}$
without ripping or tearing.

\begin{figure}
\centering
\includegraphics[scale = 0.5]{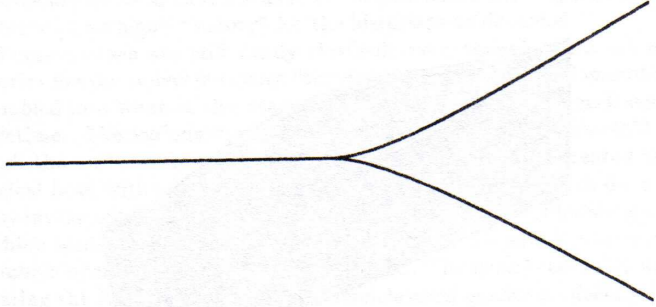}
\caption{Hausdorff but not locally Euclidean, from Visser 1996,
p253}
\end{figure}

Every topological space with a metric topology must be a Hausdorff
space. The non-Hausdorff space constructed by Visser has a
pseudo-metric rather than a metric. i.e there are distinct points,
$0_1$ and $0_2$, which are such that the distance between them
$d(0_1,0_2) =0$, violating the positive definite condition for a
metric space. The symmetry condition and triangle inequality are
preserved, so this space satisfies the conditions of a
pseudo-metric space. The fact that $0_1$ and $0_2$ are separated
by zero distance, could be used to justify the diagrammatic
representation of both these points by a single vertex to a
$Y$-shape. However, such a diagram is ambiguous at best, and
conventionally used to represent the case where the points are
identified. To obtain a metric space, with Hausdorff topology,
from a pseudo-metric space, one merely needs to identify the
points with $d(x,y) = 0$.

One can generalize these considerations to space-times which
branch globally at spacelike hypersurfaces. Taking a 4-dimensional
Lorentzian space-time $\mathcal{M}$ which can be foliated as
$\mathbb{R} \times \Sigma$, one can remove the region $[0,\infty)
\times \Sigma$, and replace it with multiple copies $[0_i,\infty)
\times \Sigma$. If the new space is obtained by merely appending
each boundary hypersurface $\{0_i \times \Sigma: i = 1,...,n\}$ to
$(-\infty,0) \times \Sigma$, the result is a non-Hausdorff space
which is locally Euclidean of dimension 4 about every point. In
contrast, identifying the corresponding points in each
hypersurface $\{0_i \times \Sigma: i = 1,...,n \}$ results in a
Hausdorff space which fails to be locally Euclidean about every
point. Points on the branching hypersurface do not have
neighbourhoods which can be homeomorphically mapped to open
subsets of $\mathbb{R}^n$. Any neighbourhood of a point on the
branching hypersurface will include a subset from each one of the
four-dimensional regions which emanate from that hypersurface.
Such a neighbourhood clearly cannot be mapped to an open subset of
$\mathbb{R}^n$ without tearing the neighbourhood in some way,
hence there cannot be a homeomorphism with an open subset of
$\mathbb{R}^n$. As a consequence, there is no tangent vector space
at any of the branch points; one cannot take the sum of vectors
which are tangent to different branches emanating from the same
point. There is therefore no tensor algebra at those points
either. The Einstein field equations, the energy conditions of
general relativity, and the expression for the local conservation
of energy-momentum, cannot hold at the branching points because
these expressions and equations are tensorial. This may be taken
as a reason to prefer the non-Hausdorff, locally Euclidean model
of branching space-time, but it is a debate which has not been
conducted in the literature.

Evaluating the many-worlds interpretation of quantum theory,
Earman draws a diagram of such a globally bifurcating space-time,
(Figure 1), and states ``I do not balk at giving up the notion,
held sacred until now, that space-time is a Hausdorff manifold.
But I do balk at trying to invent a causal mechanism by which a
measurement of the spin of an electron causes a global bifurcation
of space-time," (1986, p225). Earman's diagram of a branching
space-time is a diagram of a Hausdorff space, which fails to be
locally Euclidean at all points, so his willingness to contemplate
a non-Hausdorff space-time is rather unnecessary. Moreover, as
already suggested by Penrose, the many-worlds interpretation can
be taken to imply not a global bifurcation of space-time, but a
local branching of future light cones.

To define a space-time which branches not globally, but locally at
a point, one begins with a 4-dimensional Lorentzian manifold
$\mathcal{M}$, one removes the closure of the causal future
$J^+(x)$ of a point $x$, and then one replaces it with any number
$n$ of copies of $\overline{J^+(x)}$. One forms the union
$[M-\overline{J^+(x)}] \cup [\cup_{i=1}^n \overline{J_i^+(x)}]$.
To define a non-Hausdorff space, one can take this union, and
define its topology to be generated by the open subsets of
$[M-\overline{J^+(x)}] \cup \overline{J_i^+(x)}$, for $i=
1,...,n$. The topological space that results is non-Hausdorff, but
locally Euclidean of dimension 4 about every point, and has the
advantage, therefore, that the Einstein field equations can still
be applied at every point. Alternatively, one can define a
quotient construction which identifies the corresponding boundary
points on each copy of $\overline{J^+(x)}$. This results in a
Hausdorff topological space which fails to be locally Euclidean
about the equivalence classes which consist of more than one
point.

Both constructions can be legitimately said to represent branching
universes, but the diagrams drawn by Penrose (1979, p593) and
Visser (1996, p254) to represent these branching future light
cones correspond to the quotient constructions, not the
non-Haudorff spaces which these authors claim to be the
consequence of branching space-time.

There is certainly no guarantee that the quotient of a Hausdorff
space will itself be Hausdorff, but equally, it is false to assume
that any quotient will be non-Hausdorff. In fact, if one defines a
quotient space by the action of a group upon the original
Hausdorff space, and if the action is properly
discontinuous,\footnote{A group $G$ has a properly discontinuous
action on a set $M$ if, for any compact subset $C \subset M$, the
set $\{\phi \in G: \; \phi(C) \cap C \neq \emptyset \}$ is
finite.} then the quotient must be Hausdorff. In general, if one
treats the equivalence relationship $R$ which defines a quotient
construction on a manifold $M$ as a subset $R \subset M \times M$,
then the quotient will be Hausdorff if the projection mapping is
an open map, and if $R$ is a closed subset of $M \times M$.

As demonstrated above, examples of non-Hausdorff spaces are
usually obtained by appending extra points to existing sets, and
giving those points a neighbourhood base which is shared with a
point in the existing set. For example, take the real number line,
append a point $\star$ to it, and give $\star$ the neighbourhood
base of $0$. The union of the existing topology on $\mathbb{R}$
and the neighbourhood base of $\star$ defines a base for a
topology on $\mathbb{R} \cup \star$. In this topology one cannot
find a neighbourbood of $0$ and a neighbourhood of $\star$ which
are disjoint, hence the topology is non-Hausdorff. Sequences which
converge to $0$ also converge to $\star$ in this topology. It is
worthwhile noting, however, that this method of appending a point
can alternatively be thought of as a quotient construction; one
takes two copies of $\mathbb{R}$, and one identifies every pair of
corresponding numbers, except for the two zeroes, one of which is
re-named $\star$.

Consider Visser's assertion that in a non-Hausdorff manifold, the
dimension of the manifold doesn't necessarily equal the dimension
of the coordinate patches. This is slightly curious because it
doesn't correspond to the type of branching space-time presented
by Visser, which is non-Hausdorff but of constant local Euclidean
dimension. However, the alternative type of branching manifold,
which is Hausdorff, does correspond to the notion of a stratified
topological space, and such a space can be thought of as a
collection of manifold pieces of different dimensions.

A stratified topological space $X$ is one in which there is a
finite filtration by closed subsets

$$
X = X^n \supseteq X^{n-1} \supseteq \cdots \supseteq X^0 \supseteq
\emptyset \; ,
$$ for which the strata are the $X_i = X^i - X^{i-1}$, each of which
is a manifold. In effect, then, a stratified space is a disjoint
union of manifolds of different dimension.

A stratification should not be confused with a foliation, which is
a way of exhaustively cutting up a manifold into manifold pieces
of equal dimension. Also note that whilst every $n$-manifold, for
$n \geq 1$, will have submanifolds of smaller dimension, so that
every $n$-manifold is a stratified space, there are stratified
spaces which are not manifolds. These spaces are made up of
manifold pieces, but do not themselves constitute a manifold. The
branching hypersurfaces in an MWI space-time are 3-dimensional
topological manifolds; in the restricted topology, every point of
such a hypersurface has a neighbourhood homeomorphic with an open
subset of $\mathbb{R}^3$. With the unrestricted topology, the
points in a branching hypersurface have only neighbourhoods that
extend into the multiple branches that emanate from the
hypersurface, and this prevents the unrestricted topological space
from being locally Euclidean.

\section{Topology-change branching space-times}

The other type of branching universe, most usually discussed in
the context of general relativity and quantum gravity, is a
topology-change universe in which the number of connected
components in the future boundary is greater than the number of
components in the initial boundary. For example, if one starts
with the complex projective space $\mathbb{CP}^2$, and one removes
three copies of the interior of the closed 4-ball $D^4$, then one
obtains a 4-dimensional manifold with three boundary components
homeomorphic to $S^3$. This compact 4-manifold is of Euler
characteristic zero, hence it can be equipped with a Lorentzian
metric tensor. One can further supply this 4-manifold with a
time-orientation which renders one copy of $S^3$ as the initial
boundary, and the other two copies as the final boundary. Such a
space-time represents a universe which bifurcates in two,
(Friedman 1991, p558). Figure 5 displays a 2-dimensional analogue,
often referred to as the `trousers' space.

\begin{figure}
\centering
\includegraphics[scale = 0.3]{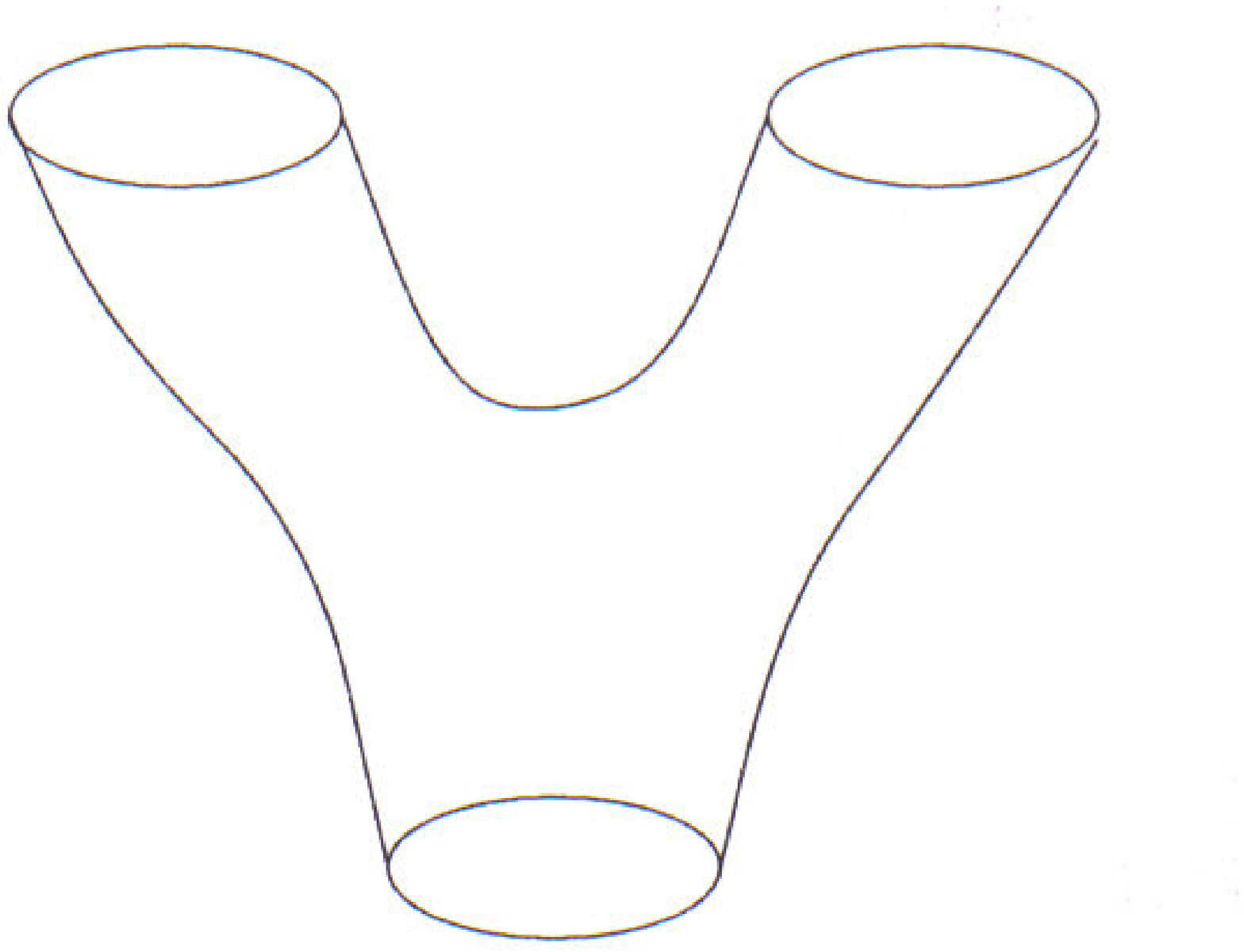}
\caption{Bifurcating topology-change space-time, from Borde
\textit{et al} 1999, p12}
\end{figure}

Each such branching space-time is a $4$-dimensional Lorentzian
manifold-with-boundary $(\mathcal{M},g)$, which interpolates
between the initial boundary $(\Sigma_i,\gamma_i,\phi_i)$ and
final boundary $(\Sigma_f,\gamma_f,\phi_f)$. The boundary of each
$\mathcal{M}$ must consist of the disjoint union of $\Sigma_i$ and
$\Sigma_f$. In addition, the restriction of the Lorentzian metric
$g$ to the boundary components must be such that $g| \Sigma_i =
\gamma_i$ and $g | \Sigma_f = \gamma_f$. Each interpolating
space-time must be equipped with a smooth matter field history
$\Phi$, which satisfies the conditions $\Phi | \Sigma_i = \phi_i$
and $\Phi | \Sigma_f = \phi_f$.

The initial $3$-manifold $\Sigma_i$ need not be homeomorphic with
the final $3$-manifold $\Sigma_f$. Hence, the transition from an
initial configuration $(\Sigma_i,\gamma_i,\phi_i)$ to a final
configuration $(\Sigma_f,\gamma_f,\phi_f)$ could be a topology
changing transition. In particular, there could be a transition to
a topology with a different number of connected components.

The notion of topology change is closely linked with the concept
of cobordism. When a pair of $n$-manifolds, $\Sigma_1$ and
$\Sigma_2$, constitute disjoint boundary components of an $n+1$
dimensional manifold, $\Sigma_1$ and $\Sigma_2$ are said to be
cobordant. Any pair of compact $3$-manifolds are cobordant,
(Lickorish 1963). Not only that, but any pair of compact
Riemannian $3$-manifolds, $(\Sigma_1,\gamma_1)$ and
$(\Sigma_2,\gamma_2)$, are `Lorentz cobordant', (Reinhart 1963).
i.e. There exists a compact $4$-dimensional Lorentzian manifold
$(\mathcal{M},g)$, with a boundary $\partial \mathcal{M}$ which is
the disjoint union of $\Sigma_1$ and $\Sigma_2$, and with a
Lorentzian metric $g$ that induces $\gamma_1$ on $\Sigma_1$, and
$\gamma_2$ on $\Sigma_2$.

This cobordism result is vital because it confirms that topology
change is possible. Even when $(\Sigma_1,\gamma_1)$ and
$(\Sigma_2,\gamma_2)$ are compact Riemannian $3$-manifolds with
different topologies, there exists an interpolating space-time.

It is impossible for a 4-dimensional manifold bounded by a couple
of non-homeomorphic 3-dimensional manifolds to be foliated by a
one-parameter family of 3-dimensional manifolds. However, it is
often said that the topology-change space-time can be described by
a Morse-function, a mapping $f:\mathcal{M} \rightarrow [0,1]$
which is such that $f| \Sigma_i = 0$ and $f| \Sigma_f = 1$. For
each $a \in [0,1]$, the set of points $f^{-1}(a)$ provides a slice
through the space-time. For a topology-change space-time there
will be at least one so-called critical slice, which contains
isolated critical points at which $\partial_\mu f = 0$. In terms
of the 2-dimensional trousers space-time, one can foliate it,
beginning at the waist, with a one-parameter family of circles
until one reaches the saddle-like branching from which the legs
emanate. There is one slice which intersects the base-point of
this saddle, and this slice comprises a pair of circles joined at
a single point. The foliation then resumes, each slice comprising
a pair of disjoint circles. The slice comprised of a pair of
circles joined at a single point is the critical slice, and the
joining point of the circles is the critical point.

By Geroch's well-known theorem (1967), if one supplies a
time-orientation that renders one hypersurface $\Sigma_i$ as the
initial hypersurface, and the other hypersurface $\Sigma_f$ as the
final hypersurface, then the topology-change space-time will
necessarily contain closed timelike curves if one adheres to the
requirement that the metric tensor must be everywhere Lorentzian
and non-degenerate. One can avoid closed timelike curves if one
permits the metric to be degenerate, and, in particular, to vanish
at isolated points, (Dowker 2003).

If one retains the requirement that the metric tensor be
well-defined and non-degenerate everywhere, then the metric tensor
will be well-defined and non-degenerate at every point of the
topology-change four-manifold $\mathcal{M}$, including the
isolated critical points of the Morse function. However, in terms
of a one-parameter slicing of $\mathcal{M}$, where each slice is
equipped with the induced 3-dimensional geometry, the
3-dimensional metric tensor is not well-defined at the critical
points. This is not the same as saying that the 4-dimensional
metric tensor is degenerate at these points. Because a critical
slice is not locally Euclidean at the critical points, and
therefore not a topological manifold, the induced 3-dimensional
metric cannot be expressed at the critical points. The critical
slices are Hausdorff, but fail to be locally Euclidean at all
points. In the simple trousers space, the point where two circles
are joined is topologically the centre of a cross, and no
cross-like neighbourhood can be mapped homeomorphically to an open
subset of $\mathbb{R}$.

This is an opposite type of failure to that suffered by the
branching MWI space-times. The latter fail to be topological
4-manifolds, but the branching hypersurfaces are topological
3-manifolds. In contrast, a topology-change space-time is a
topological 4-manifold, but it has critical slices which are not
topological 3-manifolds.

\end{document}